\documentclass{article}
\usepackage{graphicx}%
\usepackage{multirow}%
\usepackage{amsmath,amssymb,amsfonts}%
\usepackage{amsthm}%
\usepackage{mathrsfs}%
\usepackage[title]{appendix}%
\usepackage{xcolor}%
\usepackage{textcomp}%
\usepackage{booktabs}%
\usepackage{algorithm}%
\usepackage{algorithmicx}%
\usepackage{algpseudocode}%
\usepackage{listings}%
\usepackage{enumitem}  
\usepackage{float}
\usepackage{hyperref}
\usepackage[a4paper, top=1in, bottom=1in, left=0.75in, right=0.75in]{geometry}

\begin{document}

\title{GatingTree: Pathfinding Analysis of Group-Specific Effects in Cytometry Data}
\author{Masahiro Ono \\ {\footnotesize \texttt{m.ono@imperial.ac.uk}} \\ {\footnotesize Department of Life Sciences} \\ {\footnotesize Imperial College London} \\ {\footnotesize Imperial College Road, London, SW7 2AZ, United Kingdom}}

\maketitle

\begin{abstract}
Advancements in cytometry technologies have led to a remarkable increase in the number of markers that can be analyzed simultaneously, presenting significant challenges in data analysis. Traditional approaches, such as dimensional reduction techniques and computational clustering, although popular, often face reproducibility challenges due to their heavy reliance on inherent data structures. This reliance prevents the direct translation of their outputs into gating strategies for downstream experiments. Here, we propose the novel \textit{Gating Tree} methodology, a pathfinding approach that investigates the multidimensional data landscape to unravel group-specific features without the use of dimensional reduction. This method employs novel measures, including enrichment scores and gating entropy, to effectively identify group-specific features within high-dimensional cytometric datasets. Our analysis, applied to both simulated and real cytometric datasets, demonstrates that the Gating Tree not only identifies group-specific features comprehensively but also produces outputs that are immediately usable as gating strategies for pinpointing key cell populations. Furthermore, by integrating machine learning methods, including Random Forest, we have benchmarked Gating Tree against existing methods, demonstrating its superior performance. A range of supervised and unsupervised methods implemented in Gating Tree thus provides effective visualization and output data, which can be immediately used as successive gating strategies for downstream study.

\end{abstract}

\newpage

\section*{Introduction}

Cytometry plays a central role in understanding group-specific effects at the single-cell level. As cytometry technologies advance, increasing number of marker data can be analysed simultaneously, posing significant challenges in data analysis. To manage the high-dimensionality of cytometry data, manual gating alone is becoming not feasible and it is currently a common approach to employ dimensional reduction techniques along with automatic clustering methods\cite{Gassen, Fujii, Kenneth, Arvaniti}. 

Ensuring reproducibility and robustness in flow cytometric analysis is paramount, as highlighted by Cossarizza et al. \cite{Cossarizza}. Current methods, which rely heavily on the inherent structure of the dataset, face several challenges. Most notably, dimensional reduction techniques such as Principal Component Analysis (PCA) depend significantly on the dataset's intrinsic structure, resulting in new, reduced axes that amalgamate multiple markers in various ratios. This issue presents considerable challenges, even for relatively straightforward and linear techniques like PCA and its variants\cite{Ono}. Furthermore, the application of Uniform Manifold Approximation and Projection (UMAP) in cytometric data analysis can be susceptible to distortions and biases, as recognized in genomic studies \cite{Chari}. 

Moreover, popular clustering algorithms like those used in Self-Organizing Map \cite{Gassen} and K-means \cite{Fujii} introduce stochastic elements, which makes it challenging to use their outputs for downstream experiments. Further, clustering methods are sometimes followed by differential analysis to identify group-specific cell clusters within reduced dimensions \cite{Weber}, although this approach is also equally subject to the challenges imposed by dimensional reduction and computational clustering.

Efforts to mitigate these issues have included using reduced data with cluster information to generate actionable gating strategies \cite{Becht}, as well as employing cross-dataset analysis techniques \cite{Mangiola, Okada}. Nonetheless, these methods still face challenges related to the inherent data structure within each dataset. We have also extensively worked on gene expression data and flow cytometric data, using various methods for dimensional reduction and clustering \cite{Ono2013, Ono, Fujii, Bradley}, and have found significant challenges in translating data from dimensional reduction or clustering into wider research contexts.

Thus, there is a pressing need for new methodologies that can elucidate group-specific features within multidimensional marker data without relying on dimensional reduction or computational clustering with a stochastic element. Overcoming the challenges posed by the expansion of combinatorial numbers with increased marker use could provide significant benefits to the research community in academia and industry, offering new avenues for automating flow cytometric data analysis. Previous studies developed flow cytometric data analysis pipelines without the use of dimensional reduction, including CytoDx and flowGraph/SpecEnr.  CytoDx uses a generalized linear model regression using individual cell marker expression, followed by aggregating this data at the sample level to generate specific metrics \cite{CytoDx2018}. FlowGraph/SpecEnr employs a tree-based gating framework combined with t-statistics to analyze the enrichment of conditional probabilities compared to expected values within each gating strategy \cite{flowGraph2022}.

In this study, we introduce \textit{GatingTree}, a novel methodology employing a pathfinding approach to high-dimensional data, which offers immediately and directly applicable gating strategies for identifying group-specific features without the use of dimensional reduction. By developing methods for GatingTree, including Machine Learning approaches, we analyze multiple simulated and real-world datasets to demonstrate the utility of the methodology.

\subsection*{Aim of the Proposed Methodology}
The primary goal of this novel methodology is to analyze high-dimensional cytometry datasets for two-group comparisons, facilitating the identification of group-specific cell clusters compatible with downstream experimental procedures, including flow cytometric sorting. This objective is underpinned by the following key requirements:
\begin{enumerate}
    \item The methodology should generate gating strategies that can be directly applied to downstream experiments, ensuring practical applicability and seamless integration into laboratory workflows.
    \item It will eschew methods that rely heavily on the underlying data structure, such as dimensional reduction techniques, to preserve the natural variability and integrity of the data.
    \item It must efficiently handle the combinatorial complexity inherent in datasets with numerous markers, ensuring that the analysis remains computationally feasible and robust.
\end{enumerate}

\subsection*{Investigation of Multidimensional Marker Space and Construction of Gating Tree}
Our aim has led to the development of a novel approach for constructing \textit{Gating Tree}, which represents a series of gating strategies, by successively exploring the multidimensional marker space within cytometry data. This methodology employs a metaphorical Turtle to systematically navigate the landscape of the multidimensional marker space. Each location within this space, referred to as a node, is defined by a specific combination of marker states and includes cells from all samples that have the marker states. The 'height' at each node represents the degree of enrichment of cells in the experimental group relative to the control group.

The expression of markers is categorized into positive and negative, or high and low. The advantage of the simple categorisation is that it is unequivocally determined, more robust and efficient than arbitrary polygon or oval gating. This leads to each marker having three possible states, including an unassigned state (i.e., when the marker is not used), thereby structuring the space as a 3-dimensional hypercube with \( n \) dimensions. The total number of possible locations (or nodes), including the origin, is thus given by $N = 3^n$. Thus, the Turtle must still contend with exponential combinatorial expansion by selectively choosing meaningful routes based on its landscape exploration.

The Turtle's journey begins in a state of neutrality (\(0\)) across all marker axes. It may then select any marker and advance to one of two states: positive (\(+1\)) or negative (\(-1\)). Transitioning from the neutral state to either positive or negative, or from one marker state to another, constitutes a unit of movement. The Turtle's path across these markers forms a series of successive gating strategies, evolving into a branching structure, which is designated as \textit{Gating Tree}.

Each node within the Gating Tree represents a decision point at which the Turtle evaluates the landscape of the multidimensional marker space. As illustrated in Supplementary Figure S1A and S1B in Supplementary Notes, the distribution of cells along the x-axis is notably left-skewed, prompting the Turtle to commence its exploration from nodes that are directly adjacent to the origin—these nodes represent all possible single marker states, both positive and negative. The Turtle's subsequent movements are determined by the following clearly defined node rules:
\\

\noindent\fbox{%
    \parbox{\textwidth}{%
    \begin{enumerate}[label=\textbf{Rule \arabic*:}, leftmargin=*, align=left]
        \item Starting from each available marker, the Turtle investigates the landscape by moving towards directions where cells from the experimental group are significantly enriched compared to the control group. This navigational strategy is quantified and guided by the development of a new metric, the \textit{Enrichment Score}.
        \item To minimize the influence of outliers, the Turtle avoids directions that lead to a reduction in informative content. This principle of seeking paths with maximal information gain sets the foundation for the introduction of a novel metric we term \textit{Gating Entropy}, which is a variant of Conditional Entropy.
        \item The Turtle terminates exploration of any path if the resulting node contains fewer cells than a predefined minimum threshold.

    \end{enumerate}
    }
}

\begin{figure}[H]  
  \centering
  \includegraphics[width=0.85\textwidth]{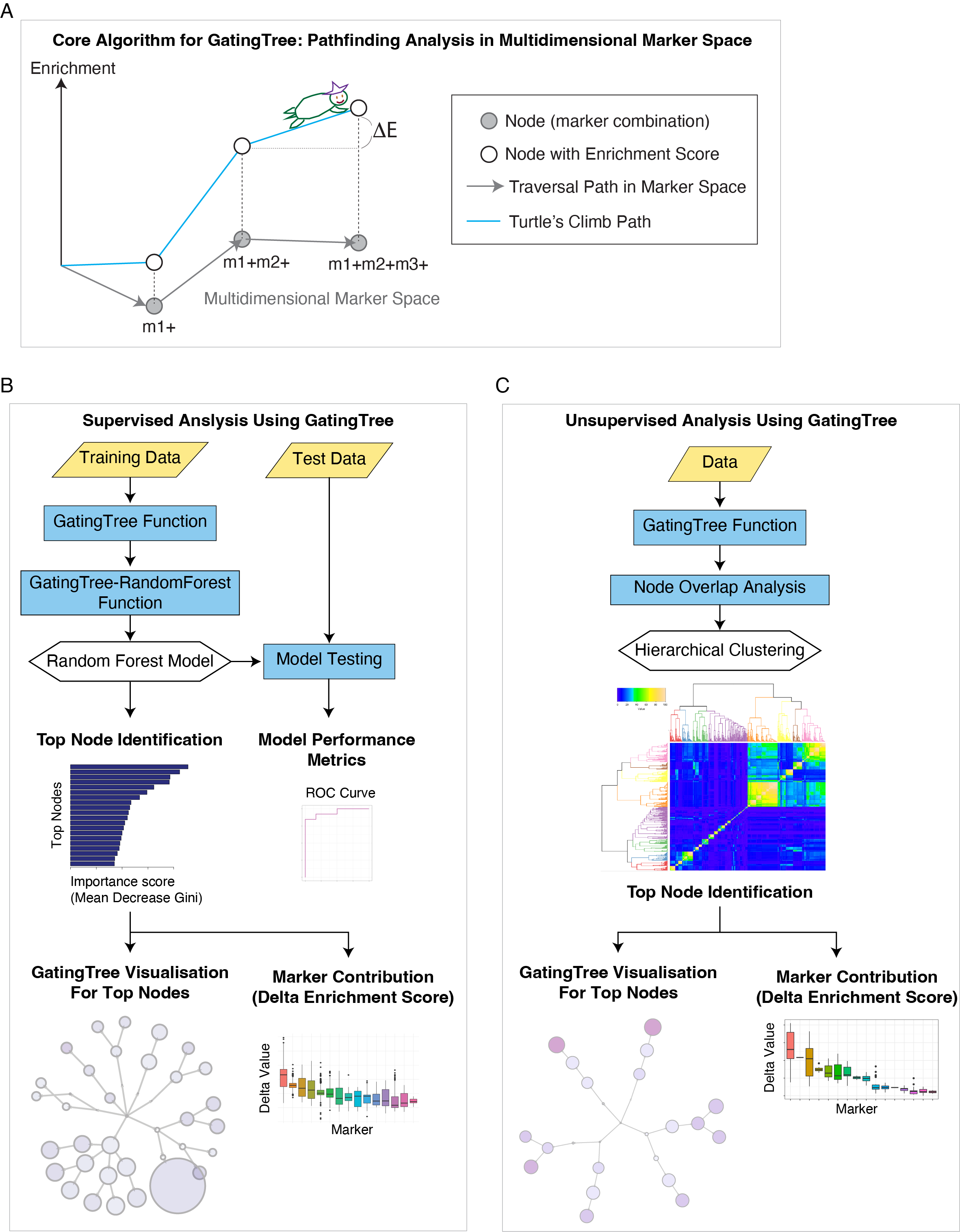}
  \caption{\textbf{Overview of the GatingTree Framework}.  
  \textbf{(A)} Schematic representation of Turtle's exploration in multidimensional marker space. The x–y plane represents combinations of flow cytometry markers, while the z-axis encodes the Enrichment Score. The Turtle symbolises a greedy pathfinding agent that strategically ascends the landscape by maximising enrichment (\(\Delta E\)) between nodes. Each node corresponds to a Boolean marker combination, and the path encodes the sequential logic of GatingTree construction.  
  \textbf{(B)} Supervised machine learning using GatingTree-derived features. Percentages of cells in Nodes are used to construct a per-sample feature matrix. A Random Forest classifier is trained to predict sample classes, and outputs include predicted probabilities, ROC curves, and feature importance scores (Mean Decrease in Gini).  
  \textbf{(C)} Unsupervised analysis of GatingTree data. Selected nodes are compared based on shared cell memberships, generating a cell-overlap matrix. Hierarchical clustering is applied to elucidate data structure, facilitating top node identification without class labels.}
  \label{fig:fig1}  
\end{figure}

Should the node rules be satisfied, the Turtle may move to multiple nodes from a single node, always 'climbing' to increase the enrichment score (Figure~\ref{fig:fig1} A). In doing so, the Turtle's paths will evolve into the successive gating strategies that identify experimental group-specific effects. This recursive application of the node rules continues until no additional nodes meet the criteria, at which point the final node in the path serves as a leaf in the Gating Tree. 

The current implementation provides a supervised machine learning method using Random Forest (Figure~\ref{fig:fig1}B) and an unsupervised analysis method based on node overlap and hierarchical clustering (Figure~\ref{fig:fig1}C).

\subsubsection*{Successive Gating Strategy Formalized as Conditional Gating}

Each marker \( m_k \) defines a basic gate determining the logical state (e.g., positive or negative) of each cell. During the pathfinding process, the Turtle incrementally constructs a gating path by adding one marker gate at a time to the existing set of gates. Specifically, a single unit move of the Turtle extends the current gate \( g_i \), which comprises the markers \(\{m_1, m_2, \ldots, m_i\}\), by incorporating the next marker \( m_{i+1} \).

This results in a new gate \( g_{i+1} \), formed by the logical AND operation between \( g_i \) and \( m_{i+1} \). Consequently, the probability of a cell passing through gate \( g_{i+1} \) is determined using the product rule of probability:
\[
P(g_{i+1}) = P(g_i \land m_{i+1}) = P(m_{i+1} | g_i) \times P(g_i)
\]

This formula underscores the conditional dependence of \( g_{i+1} \) on \( g_i \) and the sequential nature of the gating strategy employed by the Turtle.

\subsubsection*{Enrichment Score and Differential Enrichment}
To compare gating performance between an experimental group and a control group, we define the Enrichment score \( E(g_i) \) as:
\[
E(g_i) = \log\left(\frac{p_e(g_i)}{p_c(g_i)}\right)
\]
Here, \( p_e(g_i) \) and \( p_c(g_i) \) are the means of the percentages of cells that satisfy gate \( g_i \) across all samples in the experimental and control groups, respectively. 

Using the product rule of conditional probability (see Methods), the differential Enrichment score from the current gate \( g_{i} \) to the next gate \( g_{i+1} \) is defined as:
\[
\Delta E(g_{i+1}) = \log\left(\frac{p_e(g_{i+1})}{p_c(g_{i+1})}\right) - \log\left(\frac{p_e(g_i)}{p_c(g_i)}\right) = \log\left(\frac{p_e(m_{i+1}|g_{i})}{p_c(m_{i+1}|g_{i})}\right)
\]
This represents the log ratio of the conditional probabilities for the marker \( m_{i+1} \) between the experimental and control groups, given the previous gating step \( g_{i} \). By systematically investigating the distribution of $\Delta E$ for each marker given various available gates,  $\Delta E$ can be used to identify effective markers to identify the experimental-group specific effects.

\subsubsection*{Gating Entropy and Information Gain}

To mitigate the effects of outliers on the enrichment score and to assess the discriminative power of gating conditions, we introduce \textit{Gating Entropy}, a variant of conditional entropy. This metric quantifies the effectiveness of a gating condition in distinguishing between the Treatment and Control groups based on the percentages of marker-positive cells in each sample.

\paragraph{Conceptual Overview:}

At each node of the gating tree, the percentages of cells that meet the gating condition (e.g., being positive or negative for a specific marker combination) are calculated for all samples. The key idea is to evaluate whether these percentages can effectively separate the two groups. Gating entropy measures the uncertainty in correctly classifying samples into their respective groups based on the gating condition.

\paragraph{Calculation of Gating Entropy:}

1. \textbf{Classification of Samples:}

   \begin{enumerate}
   \item[1.1.] Calculate the global mean percentage \( \mu_{\text{global}} \) of marker-positive cells across all samples:

     \[
     \mu_{\text{global}} = \frac{n_{\text{Treatment}} \times \overline{p}_{\text{Treatment}} + n_{\text{Control}} \times \overline{p}_{\text{Control}}}{n_{\text{Treatment}} + n_{\text{Control}}},
     \]

where:
\begin{itemize}
    \item \( n_{\text{Treatment}} \) and \( n_{\text{Control}} \) are the numbers of samples in the Treatment and Control groups,
    \item \( \overline{p}_{\text{Treatment}} \) and \( \overline{p}_{\text{Control}} \) are the average percentages of marker-positive cells in each group.
\end{itemize}

   \item[1.2.] Classify each sample as 'High' if its percentage \( p_i \) exceeds \( \mu_{\text{global}} \), or 'Low' otherwise:

     \[
     \text{Class}_i = \begin{cases}
     \text{'High'}, & \text{if } p_i > \mu_{\text{global}}, \\
     \text{'Low'}, & \text{if } p_i \leq \mu_{\text{global}},
     \end{cases}
     \]

     where \( p_i \) is the percentage of marker-positive cells for sample \( i \).

   \end{enumerate}

2. \textbf{Computing Conditional Entropy:}

   The gating entropy \( H(\text{Group}|\text{Class}) \) is calculated using the conditional entropy formula:

   \begin{equation}
   H(\text{Group}|\text{Class}) = - \sum_{x \in \{\text{High}, \text{Low}\}} P(x) \sum_{y \in \{\text{Treatment}, \text{Control}\}} P(y|x) \log_2 P(y|x),
   \end{equation}

where:
\begin{itemize}
    \item \( P(x) \) is the proportion of samples classified as 'High' or 'Low',
    \item \( P(y|x) \) is the probability of a sample belonging to group \( y \) (Treatment or Control) given its classification \( x \) (High or Low).
\end{itemize}

3. \textbf{Interpretation:}
\begin{itemize}
   \item A gating entropy of 0 indicates perfect separation, where each classification ('High' or 'Low') exclusively contains samples from one group.
   \item An entropy closer to 1 indicates that the gating condition does not effectively distinguish between the groups.
\end{itemize}

\paragraph{Information Gain:}

Information gain quantifies the improvement in group classification when moving from one gating condition to the next in the gating tree. It is defined as the reduction in gating entropy achieved by applying an additional gating condition.

For a gating strategy \( g_{i+1} \) following the current gating \( g_i \), the information gain \( \Delta I \) is calculated as:

\[
\Delta I = H(\text{Group}|g_i) - H(\text{Group}|g_{i+1}),
\]

where:
\begin{itemize}
    \item \( H(\text{Group}|g_i) \) is the gating entropy at the current node,
    \item \( H(\text{Group}|g_{i+1}) \) is the gating entropy at the next node after applying the additional gating condition.
\end{itemize}

A higher information gain indicates that the additional gating condition significantly improves the discrimination between the Treatment and Control groups. By utilizing gating entropy and information gain, we can systematically evaluate and select gating conditions that most effectively discriminate between groups.

\subsubsection*{Algorithm for Gating Tree Construction}
The Turtle will start its journey at the origin, which becomes the Root Node. Next, the Turtle moves to all first-level child nodes, each of which represents one of all available markers, including both positive and negative states. The Turtle investigates the cell number, gating entropy, and enrichment score at each node. Then, the Turtle starts a recursive journey, in which the Turtle moves to the subsequent nodes that satisfy $\Delta E(g_{i+1}) \geq \epsilon \land \Delta I(Y \mid g_{i+1}) \geq 0 \land n_{i+1} > n_{threshold} $ (Algorithm~\ref{algo1}). Note that the Turtle always evaluates the number of cells at the next node  \( g_{i+1} \),  \( n_{i+1} \) and terminates its traversal along a given path if a node fails to meet the minimum cell number threshold.

\begin{algorithm}[!t]
\caption{Decision-Making Process of the Turtle for Navigating Nodes}\label{algo1}
\begin{algorithmic}[1]

\State \textbf{Root Rule:} Move to all available first-level child nodes.
\State \textbf{General Node Rule:}
\For{each node \( g_i \), starting from the root \( g_1 \)}
    \State Compute enrichment score \( E(g_{i}) \) and gating entropy \( H(Y | g_{i}) \)
    
    \For{each potential next node \( g_{i+1} \)}
        \State Compute \( E(g_{i+1}) \), \( H(Y | g_{i+1}) \), and cell count \( n_{i+1} \)
        \State Compute \( \Delta E(g_{i+1}) \) and \( \Delta I(Y | g_{i+1}) \)

        \If{\( \Delta E(g_{i+1}) \geq \epsilon \) and \( \Delta I(Y | g_{i+1}) \geq 0 \) and \( n_{i+1} > n_{\text{threshold}} \)}
            \State Move Turtle to node \( g_{i+1} \)
            \State Update current node to \( g_{i+1} \)
        \Else
            \State \textbf{Continue} checking other potential nodes
        \EndIf
    \EndFor
    
    \If{no valid nodes satisfy the conditions}
        \State \textbf{Terminate traversal at this path}, marking \( g_i \) as a leaf node
    \EndIf
\EndFor

\end{algorithmic}
\end{algorithm}

Specifically, if all available nodes exhibit lower enrichment scores than the current node, the latter is labeled as a 'Leaf' and also as 'Peak,' as this is crucial for understanding the geography of the data.

\begin{figure}[H]  
  \centering
  \includegraphics[width=1\textwidth]{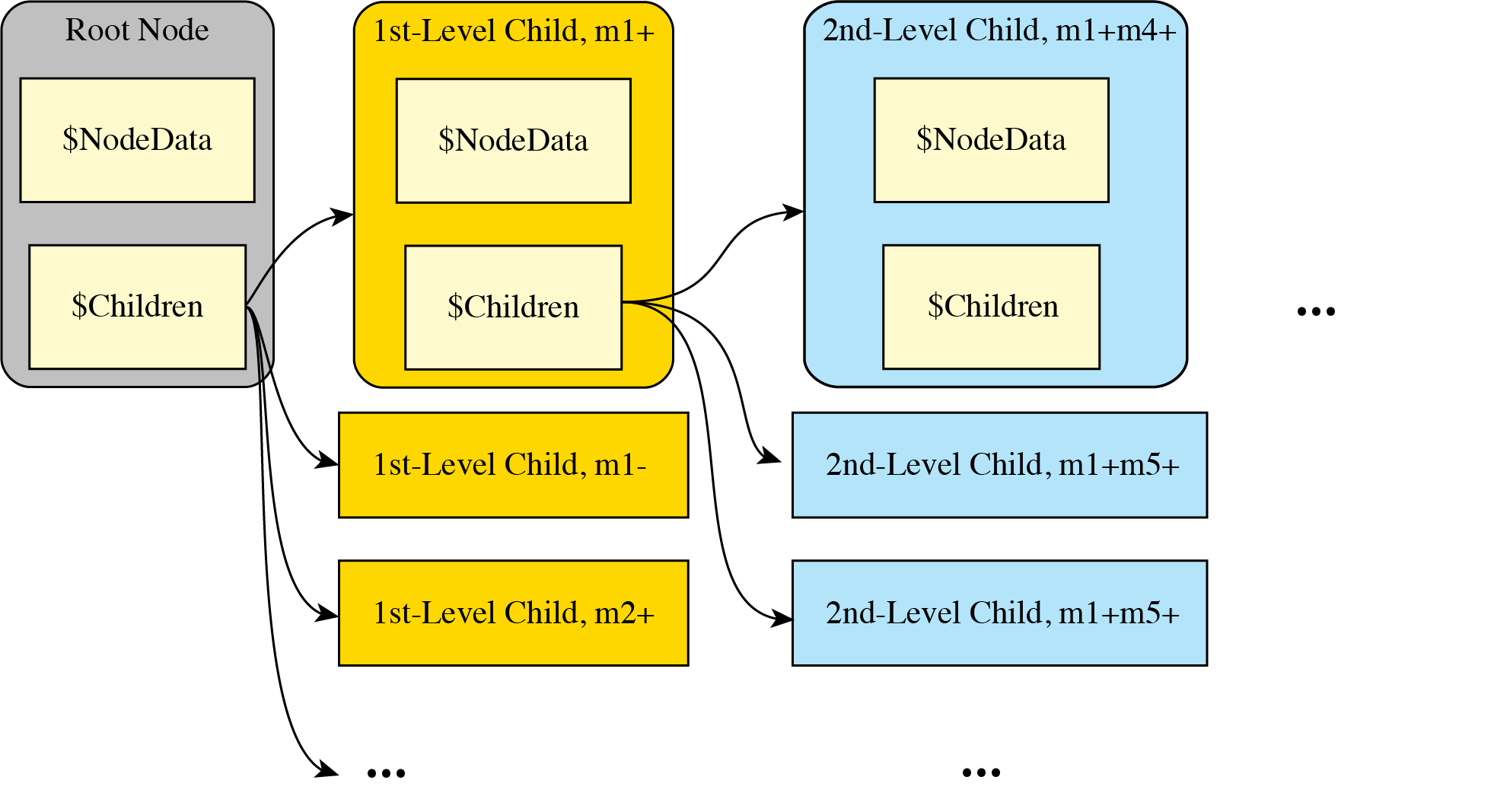}
  \caption{\textbf{Schematic Representation of the Structure of GatingTree Object.} The figure simplifies the structure of the GatingTree object, showing that each node carries the slot NodeData, which in the actual implementation is composed of multiple slots including those for enrichment score, entropy, cell indices, leaf and peak statuses, and the full path at each node, but not including expression data. Child nodes are included in the slot Children.}
  \label{fig:fig2}  
\end{figure}

All paths the Turtle traverses will constitute a tree structure, defined as a \textit{GatingTree} object (Figure~\ref{fig:fig2}). In the implementation, GatingTree is realized as a list object in the R environment. As shown in Figure~\ref{fig:fig2}, the GatingTree object maintains a tree structure, and the 'Children' slot includes direct descendant nodes. The 'NodeData' slot is implemented as multiple slots in the actual implementation, including slots for enrichment score, entropy, cell indices, the leaf status, and the full path at each node.

\subsubsection*{Pruning of Gating Tree and Node Statistics}

The complexity of the Gating Tree constructed depends on the cytometry dataset, particularly the effect size and statistical power to detect group-specific effects. In datasets with considerable differences between groups, users may wish to focus on the most salient features by removing redundancies and prioritizing more abundant cell populations over their minor subsets. To facilitate this, the Gating Tree is designed to be prunable. Users can prune the constructed Gating Tree by setting thresholds for maximum entropy, minimum enrichment score, and average cell percentage.

After pruning, nodes can be extracted from Gating Tree and subjected to statistical tests. The current implementation uses the Mann-Whitney test with p-value adjustment for multiple comparisons.

\newpage

\subsection*{Demonstrating the Efficacy and Visualization Capabilities of GatingTree with Simulated Flow Cytometric Data}

\subsubsection*{Example 1: Simulated Data Using a Mixed Gaussian Model}
\begin{figure}[H]  
  \centering
  \includegraphics[width=0.8\textwidth]{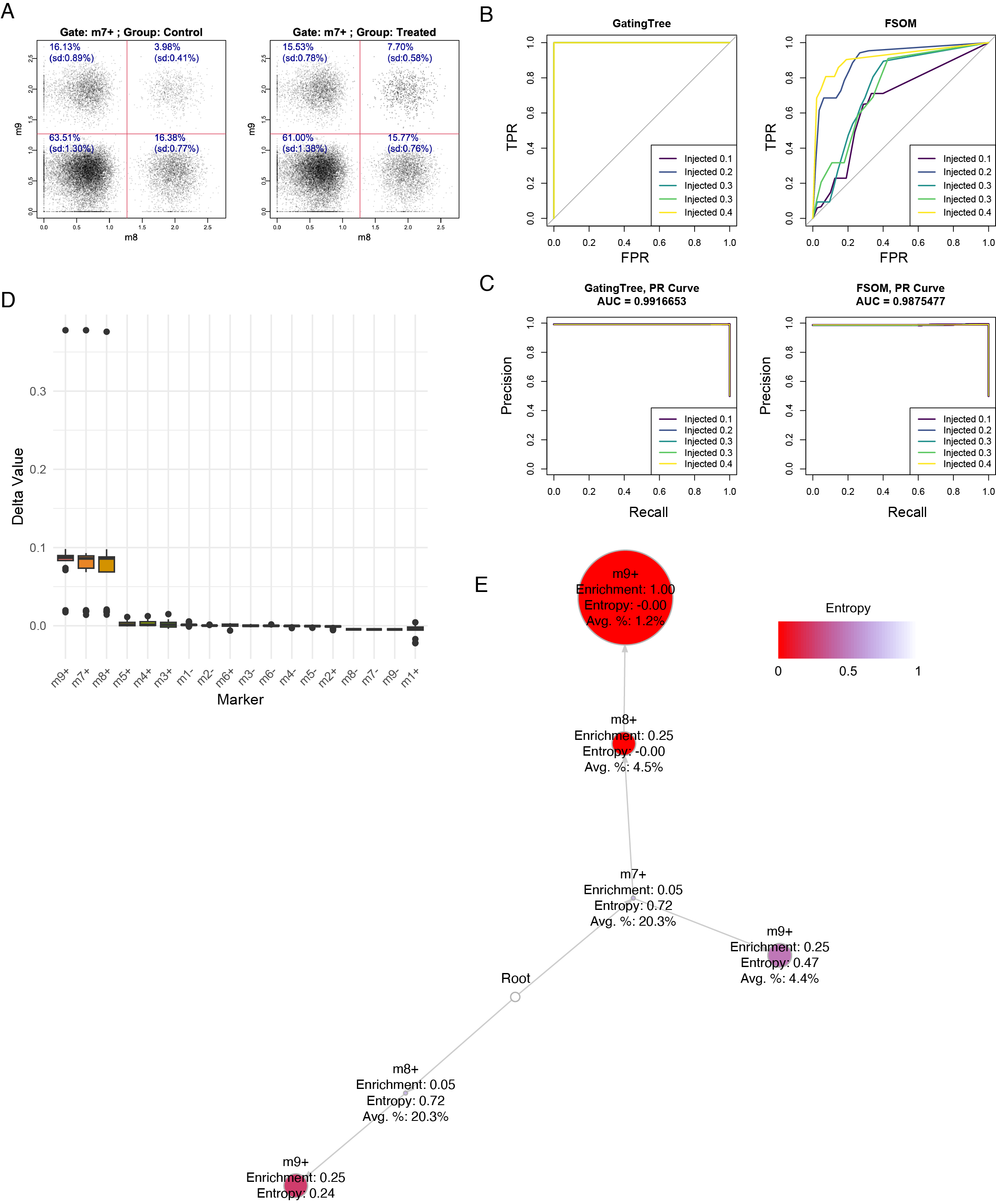}
  \caption{\textbf{GatingTree Analysis of Simulated Flow Cytometric Data using a Mixed Gaussian Model.} Nine-dimensional data were generated for two groups, each with 10 replicates. The experimental group contained a higher number of m7\textsuperscript{+}m8\textsuperscript{+}m9\textsuperscript{+} cells, randomly enhanced by one of the ratios 0.1, 0.2, 0.3, or 0.4. (A) A two-dimensional flow cytometric plot showing the increased percentage of m7\textsuperscript{+}m8\textsuperscript{+}m9\textsuperscript{+} cells in the experimental group. (B, C) ROC and Precision-Recall Curves for GatingTree and FlowSOM. (D) Boxplot of \( \Delta E \) for each marker, with an injection ratio of 0.4. (E) GatingTree plot, showing the hierarchical structure and the enrichment and entropy of each node by the size and color of each node. Avg. \% indicates the mean percentage of cells in the gate among all available cells.}
  \label{fig:fig3}  
\end{figure}

Simulated flow cytometric data were generated using a mixed Gaussian model (see Methods) and analyzed using the proposed Gating Tree method along with FlowSOM (FSOM)-based clustering. As shown in Figure~\ref{fig:fig3}A, the number of cells within the gate \(m7^{+}m8^{+}m9^{+}\) was increased by various ratios through random injections, with all cells in this gate designated as true cells. Receiver Operator Characteristic (ROC) analysis revealed that the Gating Tree method outperformed FSOM-based clustering (Figure~\ref{fig:fig3}B). The Precision-Recall Curve analysis demonstrated that both methods performed well (Figure~\ref{fig:fig3}C).

As anticipated, \( \Delta E \) values were significantly higher for the markers m7\textsuperscript, m8\textsuperscript, and m9\textsuperscript, indicating that the inclusion of each marker in the gating strategy substantially enhanced the enrichment score of the nodes (Figure~\ref{fig:fig3}D). 

Figure~\ref{fig:fig3}D shows the Gating Tree constructed by the analysis. The figure should be read as displaying successive gates from the origin (Root). The first child node, m7\textsuperscript{+}, has an enrichment of 0.05. Its child nodes, m8\textsuperscript{+} and m9\textsuperscript{+}, have the gating strategies m7\textsuperscript{+}m8\textsuperscript{+} and m7\textsuperscript{+}m9\textsuperscript{+}, respectively. The enrichment score is indicated by the size of the node, and the gating entropy is shown by the color of the node. Thus, the node m9\textsuperscript{+}, following m8\textsuperscript{+} and m7\textsuperscript{+} (i.e., m7\textsuperscript{+}m8\textsuperscript{+}m9\textsuperscript{+}), has the highest enrichment score of 1.00 and the lowest entropy, 0.0. In addition, combination gates using two of the three markers demonstrate a moderate degree of enrichment, further underscoring the capability of Gating Tree analysis to provide comprehensive insights into the dataset.

The constructed Gating Tree visualizes the successive gating strategies as a tree structure, equipped with visual representations of enrichment score and entropy, enabling users to effectively identify critical marker combinations (Figure~\ref{fig:fig3}E). 

\newpage

\subsubsection*{Example 2: A Hybrid Mass Cytometry Dataset with Simulated Cells}
Next, we analyzed a hybrid mass cytometry dataset from bone marrow cells with spiked-in leukemia cells (Acute Myeloid Leukemia, AML)  with five replicates (\textit{AML-sim}, \cite{Weber1}). Three datasets were analyzed with spiked-in cells at concentrations of 5\%, 1\%, and 0.1\%.

\begin{figure}[H]  
  \centering
  \includegraphics[width=0.95 \textwidth]{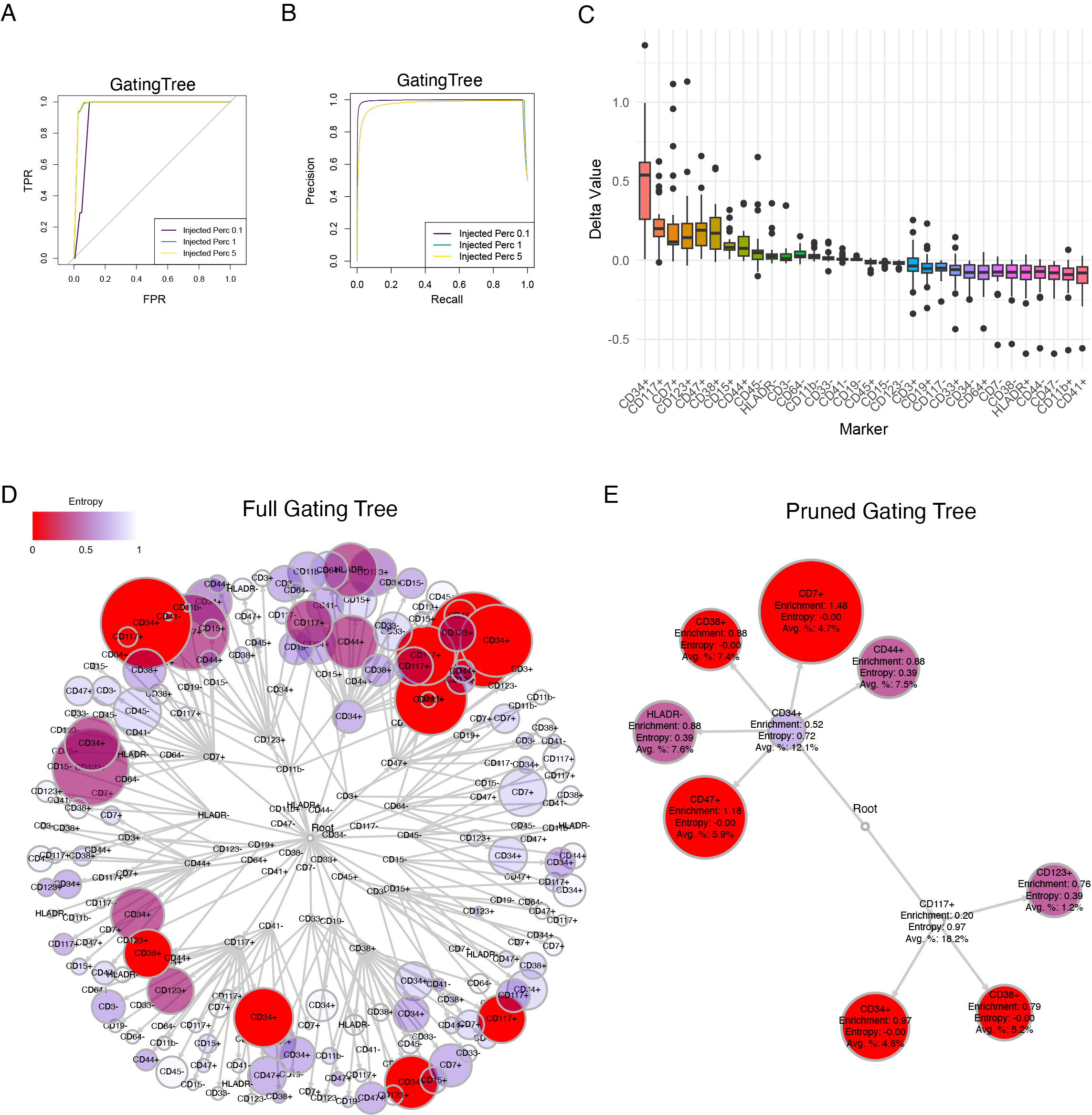}
 \caption{\textbf{Gating TreeAnalysis of Hybrid Mass Cytometry Data}. The dataset is from bone marrow cell data with spiked-in leukemic cells (\textit{AML sim}, \cite{Weber1}). has 16 marker variables and two groups with 5 replicates. Leukemic cells were injected at concentrations of 0.1\%, 1\%, and 5\% of total cells.
 (A, B) (A) ROC and (B) Precision-Recall Curves for GatingTree. (C) Boxplot of \( \Delta E \) for each marker, with an injection ratio of 5\%. (D) Gating Tree plot, showing the hierarchical structure and the enrichment and entropy of each node by the size and color of each node. Avg. \% indicates the mean percentage of cells in the gate among all available cells.(E) Pruned Gating Tree plot with the conditions, gating entropy $< 0.5$ and enrichment score $> 0.5$.} 
  \label{fig:fig4}  
 \end{figure}

ROC and Precision-Recall Curve analyses indicated that the Gating Tree method exhibited robust and high performance (Figure~\ref{fig:fig4}). $\Delta E$ analysis identified markers CD34\textsuperscript{+}, CD117\textsuperscript{+}, CD7\textsuperscript{+}, CD123\textsuperscript{+}, CD47\textsuperscript{+}, and CD38\textsuperscript{+} as having notably high values. Importantly, the expression of these markers was highly upregulated in the injected AML cells Supplementary Figure S9, which supports the effectiveness of the $\Delta E$ approach.

The dataset included 14 markers. However, all nodes showed a cell number exhaustion at the third-level children and therefore the output data contained second-level children. The construction of the GatingTree object required approximately 30 seconds. Interestingly, the first-level child nodes with a single-positive marker state—specifically those with a single-negative marker state for CD34\textsuperscript{-}, CD47\textsuperscript{-}, CD7\textsuperscript{-}, and CD38\textsuperscript{-} — were identified as a peak during the Gating Tree construction. This indicated that no additional markers improved enrichment score, suggesting that experimental group-specific cells were enriched only in nodes positivity for one of these four markers. 

To demonstrate the effectiveness of Gating Tree Pruning, Figures~\ref{fig:fig4}D and E display the Gating Tree before and after pruning, respectively. By pruning the Gating Tree under conditions where gating entropy is less than 0.5 and enrichment score is greater than 0.5, and by removing all redundant branches, the pruned Gating Tree effectively highlights the utility of key markers with high $\Delta E$ values. It identifies CD34\textsuperscript{+} as a crucial hub gate for enriching AML cells, particularly by utilizing CD7\textsuperscript{+}, CD47\textsuperscript{+}, and CD38\textsuperscript{+} markers, demonstrated high enrichment scores and low entropy values (Figure~\ref{fig:fig4}D), pinpointing the optimal marker combinations CD34\textsuperscript{+}CD7\textsuperscript{+}, CD34\textsuperscript{+}CD47\textsuperscript{+}, and CD34\textsuperscript{+}CD38\textsuperscript{+}.

\newpage

\subsection*{Runtime and Memory Analysis of GatingTree}

To effectively apply the GatingTree to real-world datasets for benchmarking purposes, we first conducted a runtime and memory analysis, which informed the analyses in subsequent sections. We used a mass cytometry dataset including 250,000 cells across 50 samples, with 33 markers (the training dataset for the CMV analysis below), to test various depths and marker numbers, as well as cell counts against runtime and memory burden.

\begin{figure}[H]
\centering \includegraphics[width=0.9 \textwidth]{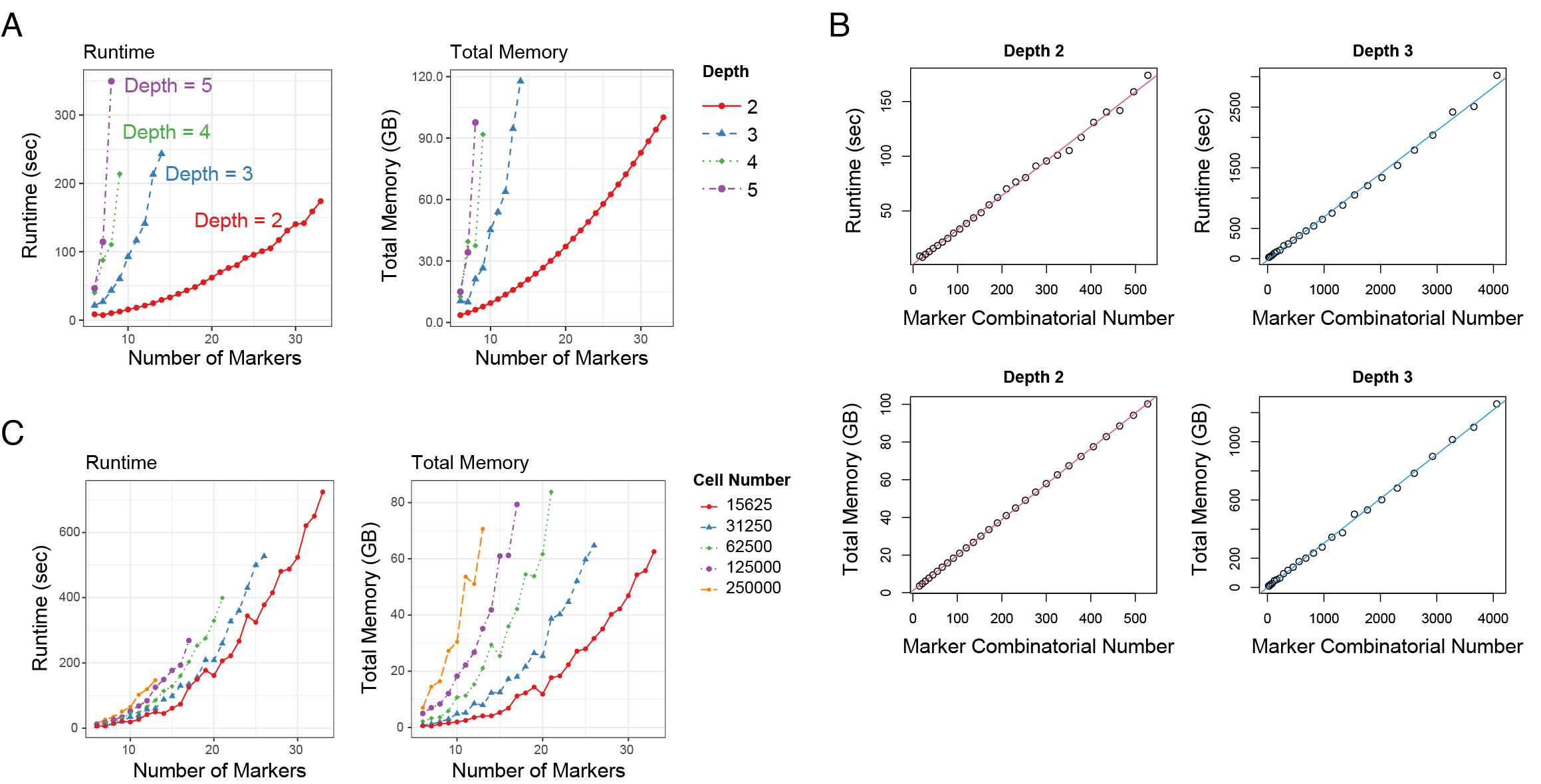} \caption{\textbf{Runtime and Memory Analysis of GatingTree}. (A) The construction of GatingTree using varying depths (2, 3, 4, and 5) and a range of marker numbers (6 to 33) was analyzed for both runtime (left) and total memory usage (right). (B) Correlations between the numbers of marker combinations and the runtime (upper panels) and total memory usage (lower panels) were examined using data from depths 2 and 3. Linear regression lines are overlayed on the graphs. All correlations demonstrated $R^2 > 0.99$, indicating very high predictability. (C) The dataset was segmented into smaller subsets to evaluate the impact of varying cell counts on the construction of GatingTree, assessing how changes in dataset size influence computational resources. } \label{fig:fig5}
\end{figure}

Using a desktop with 128 GB of memory, the analysis took up to 200 seconds for the depth-2 analysis, while a depth-5 analysis involving combinations of eight markers quickly approached 100 GB of memory usage. Therefore, using this specific dataset, extending beyond depth 3 proved challenging due to the memory burden (Figure~\ref{fig:fig5}A). Fortunately, the memory burden is predictable based on the number of marker combinations, suggesting that preliminary analysis using small marker sets can guide the analysis plan for GatingTree (Figure~\ref{fig:fig5}B). The impact of cell numbers was also significant on total memory usage, though it had less impact on runtime (Figure~\ref{fig:fig5}C). This indicates that memory usage should be analyzed and predicted for each dataset.

\newpage
 
\subsection*{GatingTreeRandomForest: GatingTree as a Supervised Method for Two-Group Classification and Node Feature Identification}

\subsubsection*{Benchmarking of the GatingTree Method Using Mass Cytometry Dataset}

\begin{figure}[H]
\centering \includegraphics[width=0.85 \textwidth]{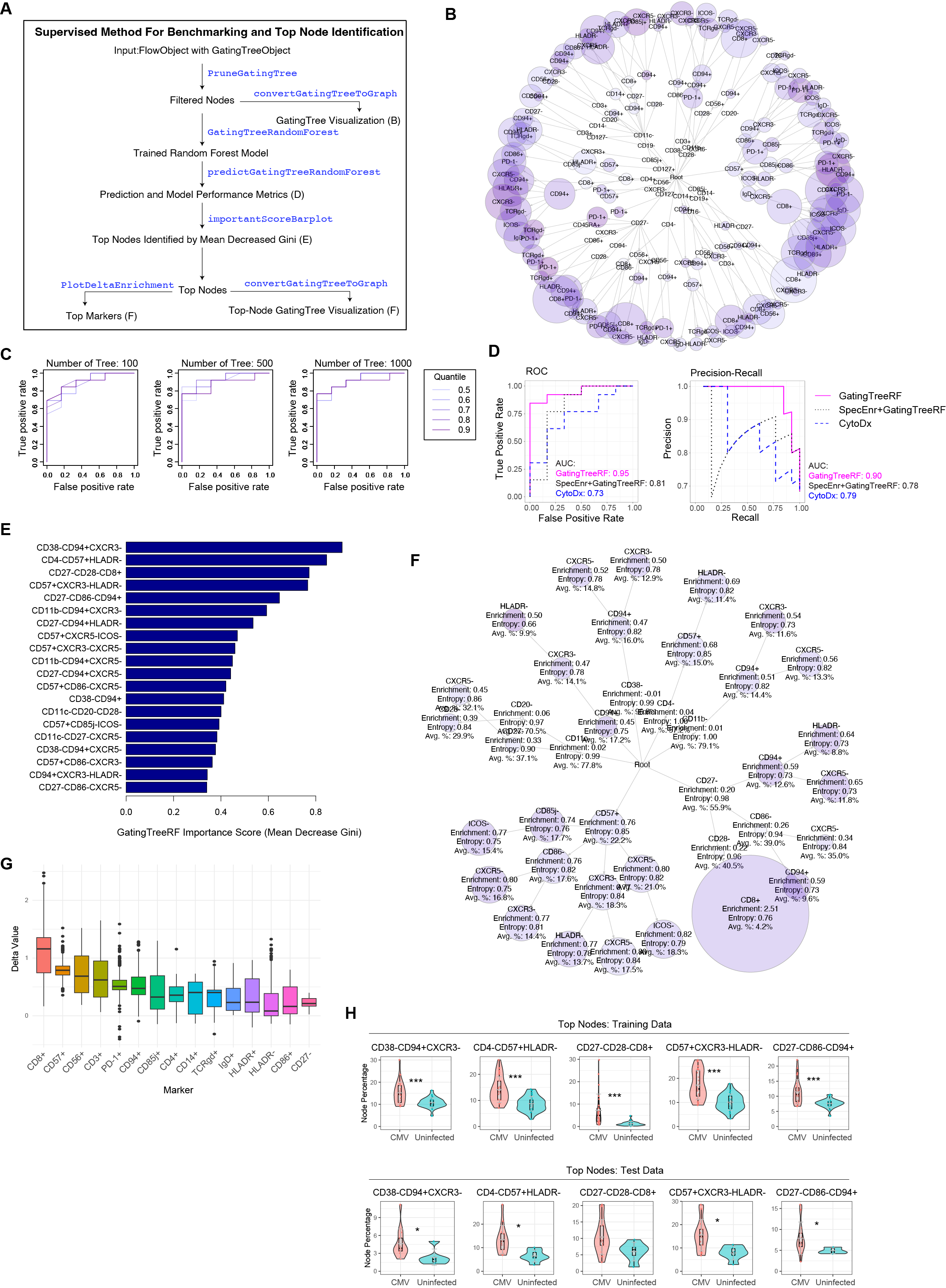} \caption{\textbf{Benchmarking of GatingTree as a Supervised Machine Learning Method.} Training and test datasets from SDY478, analyzing PBMCs from CMV-seropositive and -seronegative individuals \cite{CMV2014}. (A) Workflow for GatingTree benchmarking, with node percentage data used to train and test a Random Forest classifier. (B) Visualization of Gating Tree used for model training. (C) ROC analysis for GatingTreeRandomForest using various quantile percentages of top nodes and the indicated number of trees. (D) Comparison of GatingTreeRandomForest (GatingTreeRF), SpecEnr in combination with GatingTreeRF, and CytoDx \cite{flowGraph2022, CytoDx2018}. The number of trees used in Random Forest was 500. (E) Importance score analysis using Mean Decreased Gini to identify critical nodes. (F) Visualization of GatingTree containing top-ranked nodes only. (G) Delta Enrichment score analysis for key markers in the pruned GatingTree). (H) Violin plots illustrating cell percentages in top-ranked nodes, highlighting group-specific effects detected by the model. Statistical significance determined by the Mann-Whitney test is indicated by * for \(p < 0.05\) and *** for \(p < 0.001\)} \label{fig:fig6}
\end{figure}

To benchmark the novel GatingTree method against existing approaches, we assessed its ability to identify group-specific differences comprehensively from cytometry datasets. The GatingTree method generates node percentage data, representing the proportion of cells per sample in each node, which provides opportunities to construct a Random Forest classifier. Figure~\ref{fig:fig6}A shows an integrated pipeline implemented in the computational package for supervised two-group classification and critical node feature identification using Random Forest algorithms.

For the benchmarking purpose, training and test datasets from SDY478, which analyzed peripheral blood mononuclear cells (PBMCs) from Cytomegalovirus (CMV)-seropositive and seronegative individuals \cite{CMV2014}, pre-processed by \cite{CytoDx2018}, were used. The biology behind this dataset is that CMV seropositivity is associated with the ageing of the immune system, or immune senescence \cite{Pourgheysari2007, Mekker2012, Higdon2021}. Thus, we aimed to identify key marker combinations characteristic for CMV-seropositive individuals. 

The datasets include analyses of 33 markers across 50 training samples and 19 test samples. Given the breadth of markers, we applied the depth 3 analysis of GatingTree. By analyzing the training dataset, a Gating Tree was constructed and thresholded based on enrichment score, entropy, and node percentage for training a Random Forest model (Figure~\ref{fig:fig6}B). Receiver Operating Characteristic (ROC) analysis demonstrated the robust performance of the GatingTreeRandomForest classifier when evaluating classification accuracy across various quantile percentages of the composite score integrating the three node statistics (Figure~\ref{fig:fig6}C). Comparative benchmarking using the test dataset was performed against the existing methods, including SpecEnr and CytoDx \cite{flowGraph2022, CytoDx2018}. For this purpose, tree data was constructed using GatingTree, and significant nodes were identified by SpecEnr for further analysis with a Random Forest model (refer to Methods for details).  The benchmarking analysis showed that the GatingTreeRandomForest approach outperformed both existing methods (Figure~\ref{fig:fig6}D). The AUCs of the ROC curves were 0.90 for GatingTreeRandomForest, 0.78 for SpecEnr in combination with GatingTreeRandomForest, and 0.79 for CytoDx, respectively. Similarly, the AUCs of the Precision-Recall curves were 0.95 for GatingTreeRandomForest, 0.81 for the combined SpecEnr and GatingTreeRandomForest method, and 0.73 for CytoDx.

To identify critical nodes essential for distinguishing between groups, an importance score analysis was performed using Mean Decreased Gini of the trained GatingTree Random Forest model (Figure~\ref{fig:fig6}E). Visualization of a further pruned GatingTree, based on these top-ranked nodes by the importance score analysis, facilitated a clear and concise overview of key nodes critical to group classification (Figure~\ref{fig:fig6}F). Delta enrichment score analysis, performed on this pruned GatingTree, pinpointed key markers contributing significantly to group differentiation (Figure~\ref{fig:fig6}G). 

Violin plots of cell percentages within the identified top-ranked nodes provided by the GatingTreeRandomForest model vividly illustrate the distinct, group-specific cytometric patterns observed across both the training and test datasets (Figure~\ref{fig:fig6}H). Rather than aiming to define fixed populations, GatingTree has highlighted nodes enriched with cells expressing functional markers (Figure~\ref{fig:fig6}H).

\begin{itemize}
    \item \textbf{CD27\textsuperscript{-}CD28\textsuperscript{-}CD8\textsuperscript{+}:} CD8\textsuperscript{+} T cells without the costimulatory molecules CD28 and CD27. Previously associated with late-differentiated effector memory states and reduced proliferative capacity \cite{Pourgheysari2007, Higdon2021}.
    \item \textbf{CD57\textsuperscript{+}CXCR3\textsuperscript{-}HLA-DR\textsuperscript{-}:} Cells with CD57 expression, a marker linked to T cells and NK cells and their cytotoxicity and cellular senescence \cite{Derhovanessian2011}.
    \item \textbf{CD27\textsuperscript{-}CD86\textsuperscript{-}CD94\textsuperscript{+}, CD38\textsuperscript{-}CD94\textsuperscript{+}CXCR3\textsuperscript{-}, and CD4\textsuperscript{-}CD57\textsuperscript{-}HLA-DR\textsuperscript{-}:} Cells with the expression of CD94, linked to NK cells, without the expression of costimulatory ligands (CD86), receptors (CD27), chemokine receptors (CXCR3), or Class II MHC molecules (HLA-DR). 
\end{itemize}

\subsubsection*{Benchmarking of the GatingTree Method Using Flow Cytometry Dataset}
To determine whether the GatingTree method can effectively analyze flow cytometric data, in addition to mass cytometry data, we explored age-dependent effects on marker changes in PBMCs using the dataset SDY404. This dataset includes samples from aged (65 and older) and young (21-30) individuals~\cite{Immport}. Although the original datasets were generated to elucidate the features of high responders to vaccination, given the limited marker data and relatively low sample number, we aimed to benchmark GatingTree by classifying the two age groups while revealing age-dependent features of PBMCs. The markers analyzed include CD3 (T cell marker), CD4 (CD4\textsuperscript{+} T cell marker), CD8 (CD8\textsuperscript{+} T cell marker), HLA-DR (Class II MHC molecule), T-Bet (transcription factor for Th1 differentiation and IFN-\(\gamma\) production), CCR7 (chemokine receptor for central memory and naive T cells), EOMES (transcription factor for Th2 differentiation and T cell exhaustion), CD45RA (naive T cell marker), and Ki67 (proliferation marker). 

\begin{figure}[H]
\centering \includegraphics[width=1 \textwidth]{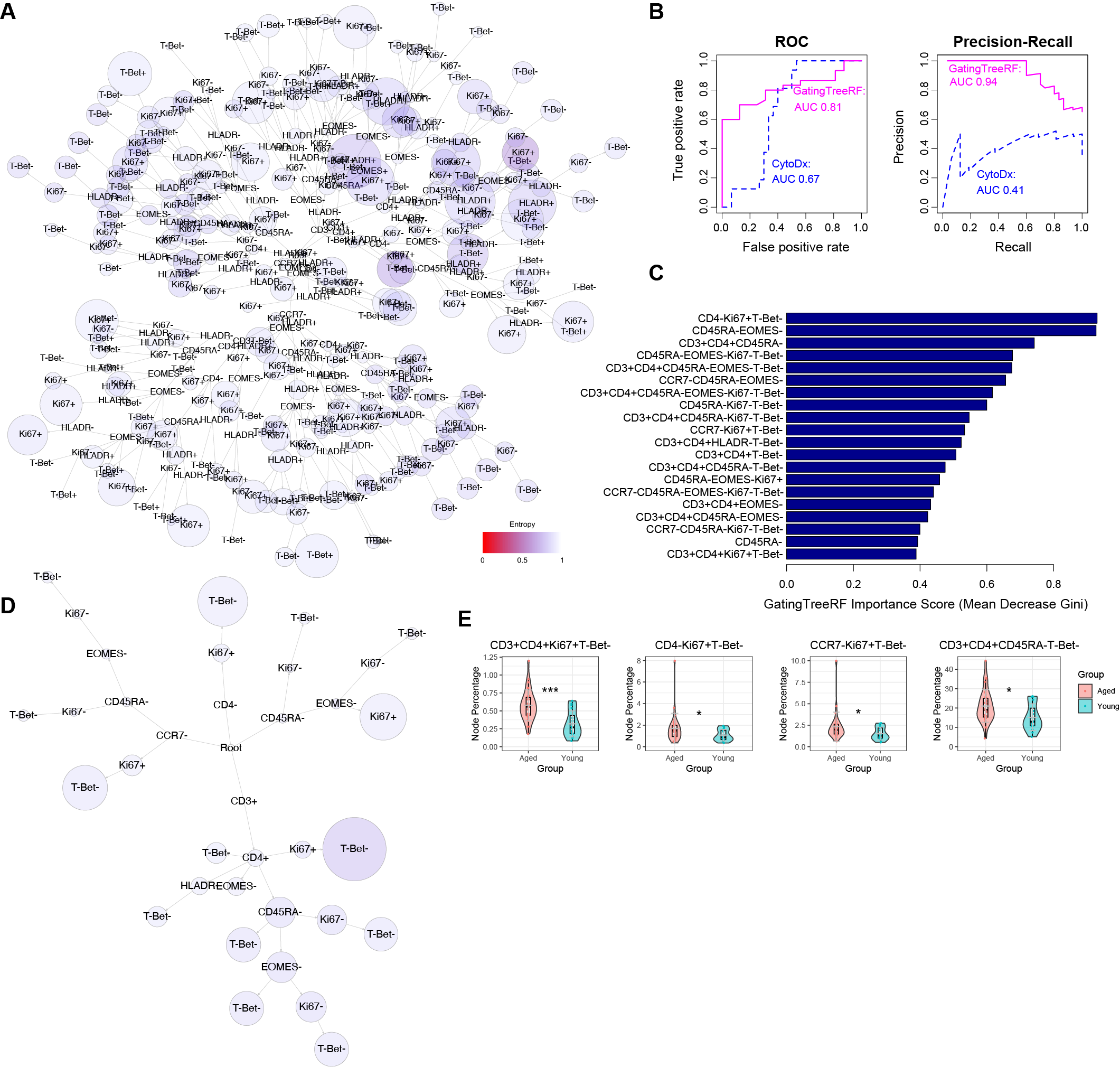} \caption{\textbf{Benchmarking of GatingTree using Flow Cytometric Data}. This analysis was conducted on the flow cytometric dataset SDY404 using the GatingTreeRandomForest approach to elucidate age-dependent effects on the marker profiles.
(A) Construction of GatingTree at depth 7, allowing for extensive marker combination analysis, pruned based on node statistics. (B) ROC and Precision-Recall analysis of GatingTreeRandomForest (GatingTreeRF) and CytoDx, using five-fold cross-validation. The Random Forest model was constructed using 1,000 trees. (C) Importance score analysis by GatingTreeRandomForest to identify nodes accounting for age-dependent features. (D) Pruned GatingTree based on the importance score analysis, highlighting critical nodes. (E) Node percentage plots of top-ranked nodes, illustrating significant differences between groups. Statistical significance determined by the Mann-Whitney test is indicated by * for \(p < 0.05\) and *** for \(p < 0.001\).
 (F) Delta Enrichment Score analysis highlighting the significance of key markers such as Ki67\textsuperscript{+}. }
\label{fig:fig7}
\end{figure}

Taking advantage of the relatively small marker panel, we constructed the GatingTree with a deep analysis at depth 7, thus enabling data exploration using up to seven marker combinations(Figure~\ref{fig:fig7}A). To effectively utilize the dataset, we employed 5-fold cross-validation to train and test the GatingTree and CytoDx models. ROC and Precision curves from aggregated cross-validation results demonstrated superior performance of the GatingTreeRandomForest model compared to CytoDx. The AUCs were 0.81 and 0.94 for GatingTreeRandomForest, versus 0.67 and 0.41 for CytoDx, respectively (Figure~\ref{fig:fig7}B).

Importance score analysis and visualization of the pruned GatingTree revealed critical nodes associated with aged individuals (Figure~\ref{fig:fig7}C-E). These nodes highlight distinct cellular behaviors, emphasizing the functional diversity within the immune system:
\begin{itemize}
    \item \textbf{CD3\textsuperscript{+}CD4\textsuperscript{+}Ki67\textsuperscript{+}Tbet\textsuperscript{-}:} CD4\textsuperscript{+} T cells expressing the proliferation marker Ki67, lacking T-bet, which suggests active proliferation without the specific Th1 lineage commitment typical for T-bet expression in CD4\textsuperscript{+} cells.
    \item \textbf{CD4\textsuperscript{-}Ki67\textsuperscript{+}Tbet\textsuperscript{-}:} Cells in a proliferative state that do not express CD4 nor T-bet.
    \item \textbf{CCR7\textsuperscript{-}Ki67\textsuperscript{+}Tbet\textsuperscript{-}:} Proliferating cells lacking the naive T cell marker CCR7 and the transcription factor T-bet.
    \item \textbf{CD3\textsuperscript{+}CD4\textsuperscript{+}CD45RA\textsuperscript{-}Tbet\textsuperscript{-}:} CD4\textsuperscript{+} T cells lacking CD45RA and T-bet, suggesting memory T cells without the evidence of Th1 polarization.
\end{itemize}
These nodes illustrate how the GatingTree approach can dissect complex immune responses by focusing on functional states rather than conventional population labels (Figure~\ref{fig:fig7}C-E).

Collectively, this benchmarking analysis confirms the utility of GatingTreeRandomForest as a supervised classification method, suitable for both flow cytometric and mass cytometry datasets.

\subsubsection*{Unsupervised GatingTree Analysis for Identifying Key Nodes Accounting for Cancer Immunotherapy Effects}

\begin{figure}[H]  
  \centering
  \includegraphics[width=0.8 \textwidth]{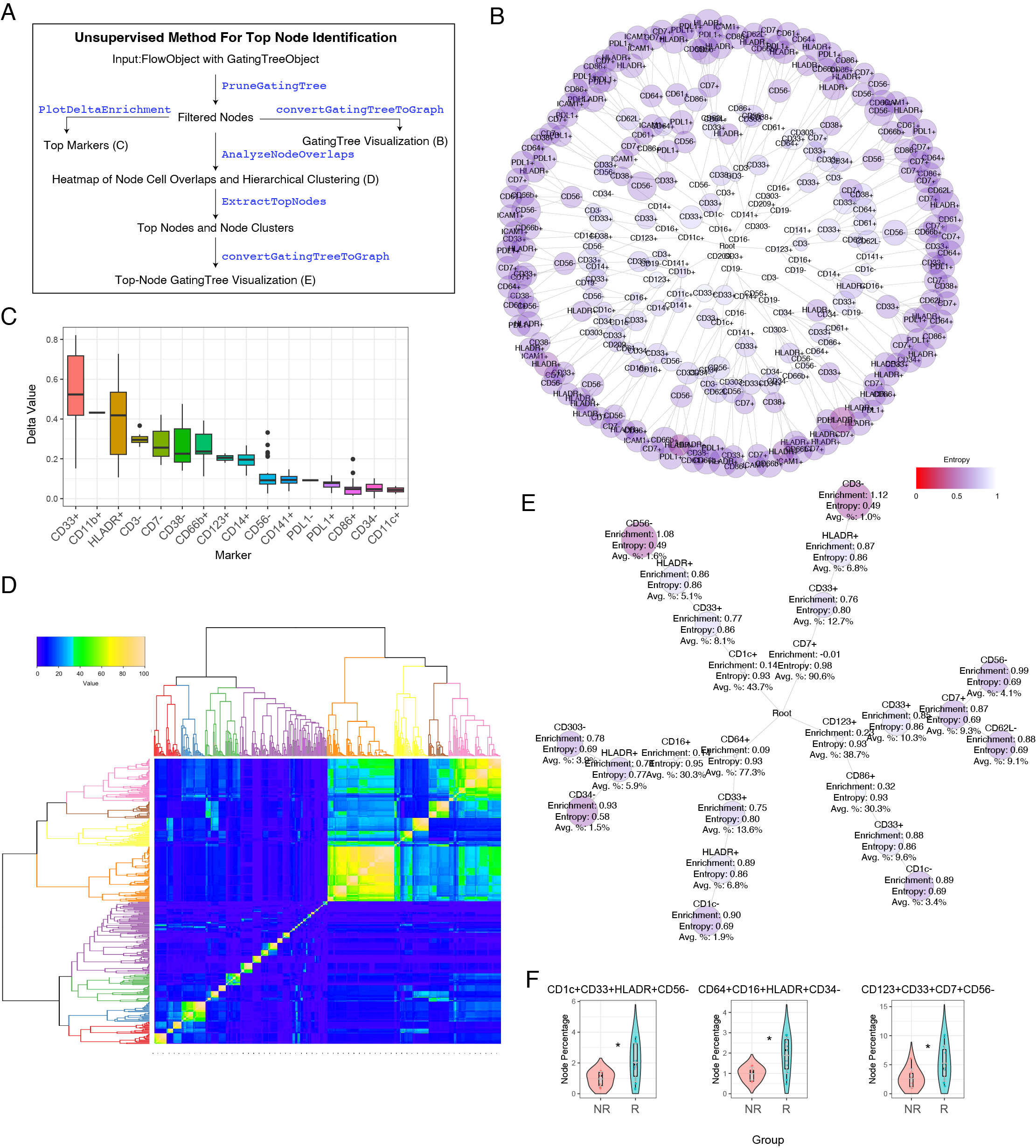}
 \caption{\textbf{Unsupervised GatingTree Analysis of Cytometry Data from Patients Under PD1 Blockade.} This analysis involves a dataset comparing responders and nonresponders to PD1 blockade therapy \cite{Krieg, Weber1}. 
(A) Workflow for the unsupervised analysis of top-ranked nodes, utilizing GatingTree with clustering analysis to handle cell overlap among nodes.
(B) Visualization of the pruned GatingTree, showing selected critical nodes.
(C) Boxplot of $\Delta E$ values illustrating differential enrichment across identified nodes.
(D) Heatmap of hierarchical clustering based on node cell overlaps, where higher values indicate greater overlap. The dendrogram's colors represent the major clusters identified.
(E) Display of the top-ranked node selected from each cluster, used to construct a succinct, pruned GatingTree.
(F) Percentage of cells in each indicated top-ranked node, comparing responders (R) to non-responders (NR) to PD-1 blockade. Statistical significance is denoted by an asterisk for $p < 0.05$ according to the Mann-Whitney test.}
     \label{fig:fig8}  
 \end{figure}

Lastly, we aimed to further develop an unsupervised analysis pipeline for GatingTree to identify critical nodes that distinguish clinically relevant groups (Figure~\ref{fig:fig8}A). The central component of this pipeline classifies GatingTree nodes into major clusters based on node overlap, defined by the percentage of shared cells among nodes. Currently, without strict control over redundancy during exploration, nodes with significant overlap may result in redundant analyses. Although the Random Forest algorithms in the supervised approach above effectively addressed this challenge, developing an unsupervised approach requires systematic resolution of this issue. To this end, we implemented hierarchical clustering of node overlap data, enabling the identification of distinct clusters of nodes.

As a demonstration of this pipeline, we analyzed a mass cytometry dataset from cancer patients undergoing immune checkpoint blockade therapy with anti-PD1 antibody, aiming to identify marker combinations distinguishing responders from nonresponders \cite{Krieg}. The dataset consisted of peripheral blood mononuclear cells (PBMCs) from 20 melanoma patients categorized clinically as responders (\(n=9\)) or nonresponders (\(n=11\)). The analysis involved 24 markers.

Given the dataset size (\(<90,000\) cells), we performed a depth-4 GatingTree analysis, identifying 64,663 non-exhausted nodes from a total of 213,052 possible nodes. A pruned GatingTree, based on node statistics, is shown in Figure~\ref{fig:fig8}B. Delta enrichment analysis identified critical contributions from markers such as CD33, CD11b, and HLA-DR (Figure~\ref{fig:fig8}C).

To effectively reduce redundancies without requiring model training and testing, hierarchical clustering analysis was applied to the node overlap data, producing a heatmap that revealed eight major clusters (Figure~\ref{fig:fig8}D). Nodes highly ranked by composite score (see Methods) represented largely independent branches of the tree and were visualized clearly in a succinct, pruned GatingTree (Figure~\ref{fig:fig8}E). Interestingly, top nodes consistently included combinations of four markers, highlighting the significance of the depth-4 analysis.

After further pruning based on gating entropy (\(<0.7\)) and enrichment score (\(>1\)), critical nodes were highlighted as shown in Figure~\ref{fig:fig8}B. Among the identified top-ranked cell populations, three innate immune subsets showed statistically significant increases in PD-1 blockade responders:
\begin{itemize}
    \item \textbf{CD1c\textsuperscript{+}CD33\textsuperscript{+}HLA-DR\textsuperscript{+}CD56\textsuperscript{-}:} Cells expressing CD1c (associated with MHC, linked to dendritic cells) and CD33 (Siglec-3, linked to myeloid cells), along with HLA-DR, but lacking CD56 (NK cell marker). This combination is indicative of dendritic cells, likely the conventional dendritic cell subtype 2 (cDC2).
    
    \item \textbf{CD64\textsuperscript{+}CD16\textsuperscript{+}HLA-DR\textsuperscript{+}CD34\textsuperscript{-}:} Cells expressing the Fc \(\gamma\) receptors CD64 and CD16, typically linked to monocytes/macrophages, with the expression of the Class II-MHC molecule HLA-DR, indicating an activated state capable of presenting antigens and mediating immune responses.
    
    \item \textbf{CD123\textsuperscript{+}CD33\textsuperscript{+}CD7\textsuperscript{+}CD56\textsuperscript{-}:} Cells expressing CD123 (the IL-3 receptor), CD33, and CD7 (associated with early hematopoietic and lymphoid development), but not expressing CD56.
\end{itemize}

\section*{Discussion}

\subsection*{Thresholding Issues and Required Data Preprocessing for GatingTree}
One of the most distinctive features of the proposed Gating Tree method is its immediate applicability in experimental settings as a tool for successive gating strategies to identify cell populations of interest. The Gating Tree employs nodes defined by simplified marker states (positive, negative, high, or low), designed to robustly facilitate the planning and execution of repeat or downstream experiments. Currently, no reliable automated methods exist for setting thresholds for marker positivity, which are typically set manually during data preprocessing. Although many experienced immunologists might prefer this manual approach, the development of automated methods could significantly benefit less experienced users and aid in achieving full automation.

Thresholds for positivity, crucial for identifying biologically meaningful cells, can be rigorously established through experimental analysis of staining controls \cite{Hulspas}. For example, while most CD4\textsuperscript{+} T cells express GITR, only those with high GITR expression (GITR\textsuperscript{high}) include cells with unique features such as Foxp3\textsuperscript{+} \cite{Ono2006}. Automating this threshold-setting process will necessitate future studies, potentially employing community-based data analysis and methods such as transfer learning \cite{Ganchev}, to establish standardized staining profiles for key antibodies and facilitate data-driven analyses.

Moreover, the current implementation of Gating Tree assumes that flow cytometric data are fully compensated or unmixed without spill-over \cite{Siddiqui}, as non-uniform signals can disrupt thresholding for marker positivity. Many high-dimensional antibody panels may not meet this ideal even after careful experimental and instrument adjustments. Promising solutions include employing new methods for modeling negative autofluorescence and making data-oriented adjustments to compensation or unmixing strategies. Alternatively, approaches such as distorted or polygon gates, which have been used in both manual and computational gating \cite{GateFinder}, could be considered, though analyzing marker-negative cells using these methods might not be straightforward. Additionally, where appropriate and possible, batch effects should be corrected using appropriate methods \cite{Combat2022} before applying the GatingTree to ensure consistent and unbiased data analysis.

\subsection*{Outstanding Computational Issues and Scalability of GatingTree}

The analysis of runtime and memory usage has illuminated the scalability challenges associated with the GatingTree method. Given the high correlation observed between the number of combinatorial markers and both runtime and memory demands (Figure~\ref{fig:fig5}), it is advisable to conduct a preliminary analysis to accurately estimate the computational resources required for each specific dataset. This initial assessment is crucial, particularly for planning extensive and deep analyses.

In real-world settings, the complexity of data is inherently constrained by practical limitations, as detailed in the section 'Constraints in Real-World Cytometry Experiments' in the Supplementary Notes. This section emphasizes that real-world constraints typically restrict the depth of analysis more significantly than might be theoretically anticipated. This has direct implications for data generation, where ensuring that a single batch of experiments can produce a sufficient amount of data for sophisticated algorithms like GatingTree is crucial.

However, despite these constraints, the expansion of data complexity remains a significant challenge. To address this effectively, our research proposes two potential solutions:

First, given that high-dimensional cytometric panels often target multiple immune cell populations simultaneously, a two-step analysis method proves to be practical. Initially, this method involves identifying major populations; subsequently, a detailed Gating Tree analysis is performed on each to fully characterize these major populations. For example, the antibody panel OMIP-069 includes 40 markers, but only 12 of these are utilized specifically to analyze CD4\textsuperscript{+} T cells, excluding markers that are used to identify the population itself \cite{OMIP069}. This two-step approach conserves computational resources by avoiding the unnecessary exploration of irrelevant markers and streamlines the GatingTree analysis. This approach can be immediately applied using the current GatingTree implementation.

Second, a shallow Gating Tree analysis combined with \(\Delta E\) and peak enrichment score analysis can provide valuable insights. Exemplary analyses at depth 3, as demonstrated in the CMV dataset (see Figure~\ref{fig:fig6}), often correlate well with deeper node analyses (Figures~\ref{fig:fig3}, \ref{fig:fig4}, \ref{fig:fig6}, and \ref{fig:fig8}). 

In the current implementation, it is encouraged to use both methods to gain insights into the data structure. Future research should focus on developing statistical methods that more effectively integrate \(\Delta E\) analysis with node exploration. These methods should aim to enhance the efficiency of exploring the high-dimensional marker space by estimating the coverage of Turtle's exploration and assessing the likelihood of identifying group-specific features.

To further enhance the method's efficiency in future implementations, it is anticipated that implementing stringent and statistically-inferred controls beyond the current non-declining enrichment score will prevent inefficiencies and reduce memory burdens. For example, introducing a cost function to selectively limit the exploration of less promising paths could enhance computational efficiency. This approach would necessitate developing a dedicated machine learning method to dynamically learn and update node exploration on the fly.

\subsection*{Integration of Machine Learning Approaches in GatingTree: Supervised and Unsupervised Techniques and their Applications}

By integrating GatingTree with supervised machine learning techniques, we have successfully benchmarked GatingTree against existing methods that do not employ dimensional reduction, such as CytoDx \cite{CytoDx2018}, to identify group-specific features. The supervised method, using Random Forest, efficiently identifies critical nodes as group-specific effects, validated through robust model performance. This performance is confirmed using both traditional training and testing with independent datasets and through cross-validation. Additionally, to benchmark GatingTree against SpecEnr, we conducted a comparative analysis using the SpecEnr algorithm and GatingTree's node statistics methods within a Random Forest model framework. Our findings indicate that GatingTree delivers superior statistical outputs. This advantage is likely attributed to the misalignment of the normality assumptions inherent in the t-test-based SpecEnr algorithm \cite{flowGraph2022} with the node percentage data generated by GatingTree. The Random Forest-based GatingTree method proves to be particularly beneficial when substantial data are available, ensuring results that are both transparent and reliable.

In our supervised GatingTreeRandomForest analysis using the CMV dataset, we identified five nodes which were increased in CMV-seropositive subjects. Among these, two nodes included cells that have previously been associated with immune ageing: the \( \text{CD27}^{-}\text{CD28}^{-}\text{CD8}^{+} \) subset, late-differentiated effector memory \( \text{CD8}^{+} \) T cells with reduced proliferative capacity \cite{Pourgheysari2007, Higdon2021}, and the \( \text{CD57}^{+}\text{CXCR3}^{-}\text{HLA-DR}^{-} \) population, where CD57 expression is linked to senescence \cite{Derhovanessian2011}. Notably, the original study reporting the CMV dataset \cite{Alpert2019} demonstrated a correlation between CMV infection status and the presence of \( \text{CD28}^{-}\text{CD8}^{+} \) T cells and mature B cells, which accelerate immune ageing. Additionally, our investigation showed that three more populations, \( \text{CD27}^{-}\text{CD86}^{-}\text{CD94}^{+} \), \( \text{CD38}^{-}\text{CD94}^{+}\text{CXCR3}^{-} \), and \( \text{CD4}^{-}\text{CD57}^{-}\text{HLA-DR}^{-} \), were also elevated in the CMV-seropositive group, potentially as newly identified effects of ageing in the immune system. While further investigations are required to establish the biological finding, these results highlight the capability of GatingTreeRandomForest to not only confirm established markers associated with CMV-induced immune senescence but also to identify new immunophenotypic signatures that merit further investigation.

While the supervised GatingTree approach is efficient and promising, it may not be suitable for all datasets, particularly those with limited sample sizes. In such instances, the unsupervised approach becomes invaluable. This approach utilizes the analysis of node cell overlaps by extensively analyzing the percentage of shared cells between pairs of nodes to identify node clusters, thereby streamlining GatingTree output. Like many unsupervised machine learning methods \cite{Geron2019}, this branch of the GatingTree method employs hierarchical clustering to classify nodes and to identify critical nodes from each cluster. This capability supplements the supervised method by enabling the identification of critical nodes within a single dataset.

In the unsupervised GatingTree analysis, the Delta enrichment analysis and the top-ranked node analysis highlighted \( \text{CD33}^{+} \) and \( \text{HLA-DR}^{+} \) as key markers for immune ageing, which resonates with findings reported by Krieg et al., who identified the \( \text{CD14}^{+}\text{CD16}^{-}\text{CD33}^{+}\text{HLA-DR}^{\text{hi}} \) population as a predictor for therapy response \cite{Krieg}.
Further biomedical research is essential to establish the broader biological and medical significance of more nuanced marker combinations identified by GatingTree. Regardless of the methodology employed, it remains crucial to biologically validate the outputs of GatingTree.

\subsection*{Extending GatingTree to Multi-Group Comparisons}
Another future direction is to extend the GatingTree method beyond two-group comparisons. Dimensional-reduction-free approaches like FloReMi employs regression analysis methods to analyze cytometry data associated with survival data \cite{FloReMi}. To extend the GatingTree method to the analysis of data without clear sample grouping information, a new strategy to explore the multidimensional marker space without being constrained by group comparison will need to be developed. This will enable the integration of node percentage data analysis using regression models, for example, such as the Cox Proportional Hazards Model and Random Survival Forests, particularly in the context of survival data. Future developments should focus on devising new node metrics that do not rely on two-group comparisons, aiming to broaden the applicability of GatingTree to more complex data scenarios.

\section*{Conclusion}

GatingTree offers a distinctive approach to data exploration without the need for dimensional reduction, generating novel and structured node proportion data ideally suited for comprehensive, data-driven analyses. This method significantly enriches successive gating strategies, proving invaluable not only in immunology but across a broad range of disciplines that utilize cytometric data. Despite existing challenges such as the need for continued development of statistical methods, enhanced computational efficiency, and better integration into extensive analytical pipelines that tackle batch effects, compensation, and unmixing corrections, GatingTree has already demonstrated substantial value. Anticipated advancements in methodology and computation are expected to further bolster the robustness and scalability of the GatingTree approach, empowering researchers to generate more detailed and reliable datasets. Such enhancements will facilitate  data-driven investigations in experimental laboratories, enabling the effective and reproducible identification of group-specific features from multidimensional marker datasets. Ultimately, GatingTree aims to transform research practices by providing deeper, data-driven insights that comprehensively address complex research questions within and beyond the realms of traditional immunology and related disciplines that use cytometry.

\section*{Methods}

\subsection*{Definition of Gates}

\textbf{Single Marker Gates:} Each marker \( m_k \) represents a basic gate that defines the logical status (e.g., positive or negative) of each cell for that marker. A single marker gate can be denoted as: \( g_k = m_k \). The probability of a cell passing through a single marker gate \( g_k \) is given by: \(P(g_k) = P(m_k)\).

\begin{flushleft}
\textbf{Combined Gates:} For sets of markers \( \{m_1, m_2, \ldots, m_i\} \), where each marker determines the logical condition of each cell, we define a combined gate \( g_{i} \) using the logical AND operation:
\end{flushleft}
\[
g_{i} = m_1 \land m_2 \land \ldots \land m_i 
\]

\begin{flushleft}
\textbf{Successive Gates:} All gates are defined to be successive and incremental in the Gating Tree method. Thus, if markers are available at the node with the gate  \( g_{i} \), the next gate  \( g_{i+1} \) is defined by the addition of another marker,  \( m_{i+1} \) 
\end{flushleft}
\[
g_{i+1} = m_1 \land m_2 \land \ldots \land m_i \land m_{i+1}
\]

\begin{flushleft}
\textbf{Probability of Successive Gates:} Given a parent gate \( g_i \) with the set \( \{m_1, m_2, \ldots, m_i\} \), the probability is calculated as:
\end{flushleft}
\[
P(g_i) = \prod_{k=1}^{i} P(m_k | g_{k-1})
\]
where $g_0$ is the origin and $p(g_0)$ is defined to be 1. Hereafter we use log probabilities, i.e.
\[
\log(P(g_i)) = \sum_{k=1}^{i} \log(P(m_k | g_{k-1}))
\]

\subsection*{Enrichment Score and Differential Enrichment}
The enrichment score \( E(g_i) \) is defined as:
\[
E(g_i) = \log\left(\frac{p_e(g_i)}{p_c(g_i)}\right)
\]

The differential Enrichment score \( \Delta E(g_i) \) is defined as:

\[
\Delta E(g_i) = E(g_i) - E(g_{i-1}) = \log\left(\frac{p_e(g_i)}{p_c(g_i)}\right) - \log\left(\frac{p_e(g_{i-1})}{p_c(g_{i-1})}\right)
\]

Using the product rule of conditional probability:
\[
\Delta E(g_i) = \log\left(\frac{p_e(m_i|g_{i-1}) \times p_e(g_{i-1}) \cdot p_c(g_{i-1})}{p_c(m_i|g_{i-1}) \times p_c(g_{i-1}) \cdot p_e(g_{i-1})}\right) = \log\left(\frac{p_e(m_i|g_{i-1})}{p_c(m_i|g_{i-1})}\right)
\]

Thus, the differential Enrichment score between successive gating steps \( g_i \) and \( g_{i-1} \) is given by:
\[
\Delta E(g_i) = \log\left(\frac{p_e(m_i|g_{i-1})}{p_c(m_i|g_{i-1})}\right)
\]

Note that the calculation of \( \Delta E \) employs a base-2 logarithm (log2).

\subsection*{Gating Entropy and Information Gain}

Gating entropy is a measure derived from conditional entropy that quantifies the effectiveness of a gating condition in distinguishing between two groups of samples (e.g., Treatment and Control). It assesses how well the gating condition separates the samples based on the percentage of cells meeting certain criteria within each sample.

\paragraph{Overview:}

Each sample contains a percentage of cells that meet the gating condition (e.g., being positive for a specific marker). The goal is to determine whether this percentage can effectively distinguish between the two groups of samples. A lower gating entropy indicates a better separation, with an entropy of 0 representing a perfect distinction and an entropy approaching 1 indicating no discriminative power.

\paragraph{Calculation Steps:}

1. \textbf{Compute Group Averages:}

   For each group (Treatment and Control), calculate the average gated percentage across all samples in that group:

   \[
   \overline{p}_\text{Treatment} = \frac{1}{n_\text{Treatment}} \sum_{i=1}^{n_\text{Treatment}} p_i^\text{Treatment}, \quad
   \overline{p}_\text{Control} = \frac{1}{n_\text{Control}} \sum_{i=1}^{n_\text{Control}} p_i^\text{Control},
   \]

   where \( p_i^\text{Group} \) is the gated percentage for sample \( i \) in the specified group, and \( n_\text{Group} \) is the number of samples in that group.

2. \textbf{Determine the Global Mean Percentage:}

   Calculate the global mean of the group averages:

   \[
   \mu_{\text{global}} = \frac{n_\text{Treatment} \times \overline{p}_\text{Treatment} + n_\text{Control} \times \overline{p}_\text{Control}}{n_\text{Treatment} + n_\text{Control}}.
   \]

3. \textbf{Classify Samples Based on the Global Mean:}

   For each sample, classify it as 'High' if its gated percentage exceeds the global mean, or 'Low' otherwise:

   \[
   \text{Class}_i = \begin{cases}
   \text{'High'} & \text{if } p_i > \mu_{\text{global}}, \\
   \text{'Low'} & \text{if } p_i \leq \mu_{\text{global}}.
   \end{cases}
   \]

4. \textbf{Construct the Contingency Table:}

   Create a contingency table that cross-tabulates the sample classifications ('High' or 'Low') against the actual group labels (Treatment or Control):

   \begin{center}
   \begin{tabular}{c|cc}
   \hline
   & Treatment & Control \\
   \hline
   High & $n_{\text{High,Treatment}}$ & $n_{\text{High,Control}}$ \\
   Low & $n_{\text{Low,Treatment}}$ & $n_{\text{Low,Control}}$ \\
   \hline
   \end{tabular}
   \end{center}

5. \textbf{Calculate Conditional Probabilities:}

   For each classification ('High' or 'Low'), compute the probabilities of being in each group:

   \[
   p(\text{Group} | \text{Class}) = \frac{n_{\text{Class,Group}}}{n_{\text{Class}}},
   \]

   where \( n_{\text{Class}} \) is the total number of samples classified as 'High' or 'Low'.

6. \textbf{Compute the Gating Entropy:}

   The gating entropy \( H(\text{Group} | \text{Class}) \) is calculated using the conditional entropy formula:

   \[
   H(\text{Group} | \text{Class}) = -\sum_{\text{Class} \in \{\text{High}, \text{Low}\}} p(\text{Class}) \sum_{\text{Group} \in \{\text{Treatment}, \text{Control}\}} p(\text{Group} | \text{Class}) \log_2 p(\text{Group} | \text{Class}),
   \]

   where \( p(\text{Class}) = \frac{n_{\text{Class}}}{n_{\text{Total}}} \) is the proportion of samples in each classification.
   
7. \textbf{Baseline Entropy at the root node} \( H(Y) \), without any gating condition, is defined as follows:
\[
H(Y) = - \sum_{k \in \{\text{Treatment}, \text{Control}\}} p(k) \log_2 p(k)
\]

\paragraph{Interpretation:}

- An entropy of 0 indicates perfect discrimination, where each classification ('High' or 'Low') contains samples from only one group.
- An entropy close to 1 suggests that the gating condition does not distinguish between the groups, with the classifications containing an even mix of Treatment and Control samples.

\paragraph{Example:}

Suppose we have the following data:

- Treatment group: 3 samples with gated percentages of 80\%, 85\%, and 90\%.
- Control group: 3 samples with gated percentages of 10\%, 15\%, and 20\%.

\textbf{Calculation:}

1. Compute group averages:

   \[
   \overline{p}_\text{Treatment} = \frac{80 + 85 + 90}{3} = 85\%, \quad
   \overline{p}_\text{Control} = \frac{10 + 15 + 20}{3} = 15\%.
   \]

2. Determine the global mean:

   \[
   \mu_{\text{global}} = \frac{3 \times 85\% + 3 \times 15\%}{6} = 50\%.
   \]

3. Classify samples:

   - Samples with percentages above 50\% are classified as 'High'.
   - Treatment samples: all 'High'.
   - Control samples: all 'Low'.

4. Contingency table:

   \begin{center}
   \begin{tabular}{c|cc}
   \hline
   & Treatment & Control \\
   \hline
   High & 3 & 0 \\
   Low & 0 & 3 \\
   \hline
   \end{tabular}
   \end{center}

5. Calculate probabilities:

   - \( p(\text{High}) = \frac{3}{6} = 0.5 \), \( p(\text{Treatment} | \text{High}) = 1 \), \( p(\text{Control} | \text{High}) = 0 \).
   - \( p(\text{Low}) = \frac{3}{6} = 0.5 \), \( p(\text{Treatment} | \text{Low}) = 0 \), \( p(\text{Control} | \text{Low}) = 1 \).

6. Compute gating entropy:

   \[
   H(\text{Group} | \text{Class}) = -[0.5 \times (1 \times \log_2 1 + 0 \times \log_2 0) + 0.5 \times (0 \times \log_2 0 + 1 \times \log_2 1)] = 0.
   \]

This result indicates perfect discrimination between the Treatment and Control groups using the gating condition.

\subsection*{Datasets}
The \textit{AML-sim} and \textit{PD-1} datasets were from the R package HDCytoData \cite{Weber1}. The \textit{CMV} and \textit{Ageing} datasets were originally generated by the studies SDY478 \cite{CMV2014} and SDY404 \cite{Immport}. A mixed Gaussian data was generated by sampling autofluorescence cells from $\mathcal{N}(1000, 100^2)$ and marker positive cells from $\mathcal{N}(2000, 200^2)$ for nine markers independently and randomly. The final dataset $\mathbf{D}$ is obtained by applying logarithmic transformation. Subsequently, the proportion of target cell cluster, \(m8^{+} m9^{+} m10^{+}\) cells within each data was identified, randomly reassigned within the dataset to simulate additional cell factor influences (injected cells).

\subsection*{Computational Environment}
The computational analyses were conducted on an Apple M2 Ultra system running macOS version 14.7.4. The machine was equipped with 128 GB of RAM and 2 TB of total storage. The data processing and analyses were performed using R version 4.4.1.

\section*{Implementation}
Methods for constructing GatingTree are implemented as an R package using base R functions and the R package \textit{dplyr}, using a special S4 object system, \texttt{FlowObject}, which facilitates a comprehensive workflow encompassing data initialization, preprocessing, execution of GatingTree analyses, and data visualization. 

\subsubsection*{GatingTree Visualization}
The Tree visualization is realised by importing functions from the CRAN packages \textit{DiagrammeR}\cite{Iannone} and \textit{data.tree} \cite{Glur}.

\subsubsection*{Random Forest Modelling and Benchmarking}

The Random Forest algorithm implemented in GatingTreeRandomForest utilized the \texttt{randomForest} function from the R package \textit{randomForest} \cite{randomForest}. The number of variables randomly sampled as candidates at each split was set to the square root of the total number of variables in the dataset. Cross-validation was conducted by splitting the data using the \texttt{createFolds} function from the R package \textit{caret} \cite{caret}. Importance score analysis was performed using the Mean Decreased Gini values of the variables from the Random Forest model.

Benchmarking was performed against the existing methods including SpecEnr, CytoDx \cite{flowGraph2022, CytoDx2018}, and FlowSOM \cite{Gassen}. ROC and Precision-Recall analyses were conducted using the R packages \textit{pROC} \cite{Robin} and \textit{PRROC} \cite{Keilwagen}. 

\subsubsection*{Pruning of Gating Tree}
To prune the Gating Tree, the Gating Tree object is traversed to collect the enrichment score, entropy, and the average percentage of cells relative to the total cells for all nodes, resulting in a data frame. Thresholds are then established to filter out nodes with low enrichment scores, high entropy, or an insufficient number of cells. Nodes in the Gating Tree are subsequently labeled as either to be pruned (\texttt{Prune = TRUE}) or retained (\texttt{Prune = FALSE}), which labeling is used to prune the tree.

For GatingTreeRandomForest analysis, the filtration of nodes was performed using the quantile thresholding of the composite score integrating the three node statistics using the following formula, selecting the top ranked nodes above the percentile threshold specified. For each node $n$,

\begin{equation}
s = Z_{\text{Enrichment}}(n) + Z_{\text{Entropy}}(n) + Z_{\text{Average\_proportion}}(n)
\end{equation}
where:
\begin{align*}
Z_{\text{Enrichment}}(n) &= \frac{\text{Enrichment}(n) - \text{mean}(\text{Total Enrichment})}{\text{sd}(\text{Total Enrichment})}, \\
Z_{\text{Entropy}}(n) &= \frac{10^{-\text{Entropy}(n)} - \text{mean}(10^{-\text{Total Entropy}})}{\text{sd}(10^{-\text{Total Entropy}})}, \\
Z_{\text{Average\_proportion}}(n) &= \frac{\text{Average\_proportion}(n) - \text{mean}(\text{Total Average\_proportion})}{\text{sd}(\text{Total Average\_proportion})}.
\end{align*}

\subsubsection*{Node Percentage Analysis and Visualization}
After pruning the Gating Tree, statistical tests are employed to validate the remaining nodes. Given that node percentage data typically do not follow a normal distribution, non-parametric tests are more appropriate. Accordingly, the current implementation utilizes the Mann-Whitney U test to compare node distributions between groups. To account for multiple testing, p-values are adjusted using the Benjamini-Yekutieli (BY) method to control the false discovery rate under conditions of dependency among tests \cite{Benjamini}. Violin plots for node percentage data were generated using the R packages ggplot2 \cite{ggplot2} and gridExtra \cite{gridExtra}.

\section*{Workflow of the GatingTree Package Using the S4 Object System}

\subsection*{Data Structure}
\begin{figure}[H]  
  \centering
  \includegraphics[width=0.7\textwidth]{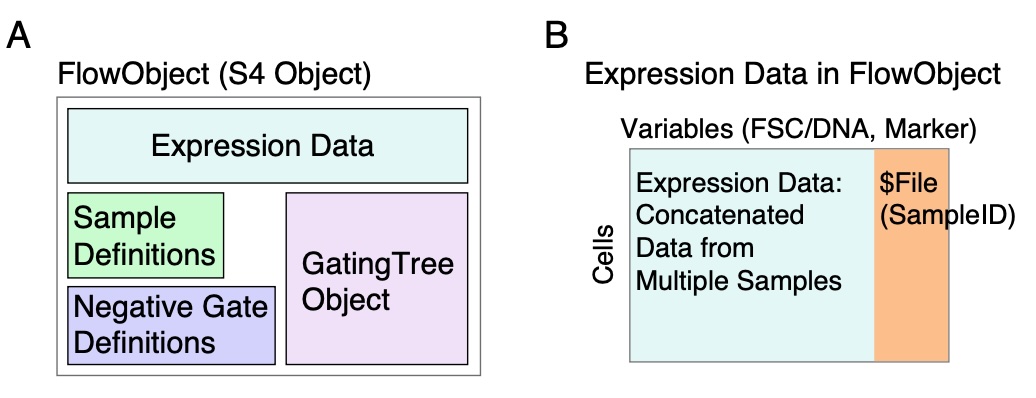}
  \caption{Structure of \texttt{FlowObject} and its components. (A) Slots in \texttt{FlowObject} include Expression Data, Sample Definitions, which define sample grouping, Negative Gate Definitions, and GatingTree Object. (B) Expression Data in \texttt{FlowObject} is created by concatenating expression data from all samples. The column `File` identifies sample origins.}
  \label{fig:fig9}  
\end{figure}

All data are stored and systematically analyzed within \texttt{FlowObject} in the GatingTree package (Figure~\ref{fig:fig9}A and B).

\subsection*{Overview of Workflow}
Immediately after initialization by \texttt{CreateFlowObject}, \texttt{FlowObject} includes expression data and sample definitions. This is followed by data transformation, employing a logarithmic transformation. \texttt{DefineNegatives} allows users to define negative gates. Subsequently, the GatingTree package offers an option to moderate extreme negative values by enforcing a normal distribution within the negative (autofluorescence) range (\texttt{NormAF}). This adjustment facilitates effective visualization of histograms and 2D plots, while preserving the quantitative accuracy of logarithmic transformation for positive cells.

\begin{figure}[H]  
  \centering
  \includegraphics[width=0.5\textwidth]{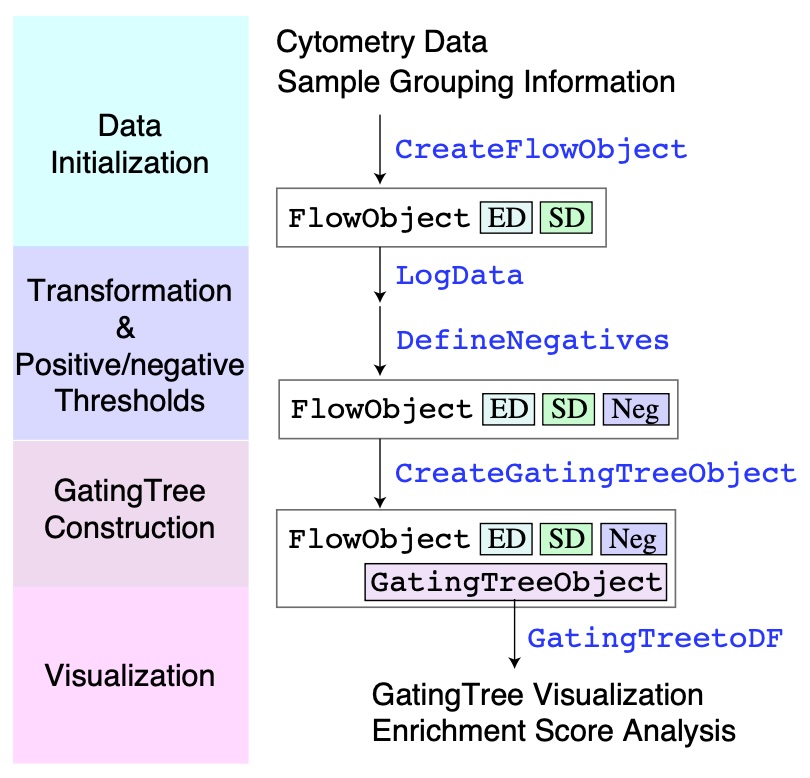}
  \caption{Overview of the workflow depicting the significance of each function at each step of data analysis, clarifying the sequence of function applications. ED: expression data; SD: sample definitions; Neg: negative gate definitions.}
  \label{fig:fig10}  
\end{figure}

Upon completing data initialization, transformation, and negative threshold definitions for all markers, the function \texttt{CreateGatingTreeObject} can be applied to \texttt{FlowObject}, creating a \texttt{GatingTree object}, which is a list object, within \texttt{FlowObject} (Figure~\ref{fig:fig10}). This will be followed by visualization of the constructed GatingTree and enrichment score analysis to determine the impact of each marker state on the enrichment score.

For a detailed outline of the current analysis workflow, including sample codes and outputs, please refer to the Supplementary Note accompanying this paper. This note provides comprehensive backgrounds to the nature of cytometry data and the specific analysis techniques implemented in the GatingTree package.

\subsection*{Acknowledgements}
MO was supported by a CRUK Programme Foundation Award (DCRPGF/100007).

\subsection*{Conflict of interest}
The author declares no conflict of interest.

\subsection*{Code availability} 
The source code for the GatingTree software, implemented as an R package, is freely available through the MonoTockyLab GitHub repository (\url{https://github.com/MonoTockyLab/GatingTree}). Comprehensive vignettes and manual pages are accessible at \url{https://monotockylab.github.io/GatingTree/}.

\bibliographystyle{unsrt}
\bibliography{GatingTree_CytometryA2025Revision}

\end{document}